\documentclass[10pt,letterpaper,twocolumn]{article} 

\usepackage{ol2}

\usepackage{graphicx}
\usepackage{dcolumn}
\usepackage{bm}
\usepackage{amsxtra}
\usepackage[latin9]{inputenc}
\usepackage[small]{caption}

\usepackage{epstopdf} 
\usepackage[colorlinks=true, citecolor = blue, linkcolor = blue, pdfborder={0 0 0}]{hyperref}

\begin{document}

\twocolumn[ 
\title{Narrow-line magneto-optical trap for dysprosium atoms}
\author{T. Maier$^{1}$, H. Kadau$^{1}$, M. Schmitt$^{1}$, A. Griesmaier$^{1}$ and T. Pfau$^{1,*}$}
 \address{
 $^1$5. Physikalisches Institut, Universit\"{a}t Stuttgart, Pfaffenwaldring 57, 70569 Stuttgart, Germany \\
 $^*$Corresponding author: t.pfau@physik.uni-stuttgart.de}
%
\begin{abstract}
We present our technique to create a magneto-optical trap for dysprosium atoms using the narrow-line cooling transition at 626$\,$nm to achieve suitable conditions for direct loading into an optical dipole trap. The magneto-optical trap is loaded from an atomic beam via a Zeeman slower using the strongest atomic transition at 421$\,$nm. With this combination of two cooling transitions we can trap up to $2.0\cdot10^8$ atoms at temperatures down to 6$\, \mu$K. This cooling approach is simpler than present work with ultracold dysprosium and provides similar starting conditions for a transfer to an optical dipole trap.
\end{abstract}


 ]



Strongly dipolar quantum gases enable the observation of many-body phenomena with a balanced interplay between the isotropic, short-range contact interaction and the anisotropic, long-range dipolar interaction \cite{Review}. Particular phenomena for a bosonic system are rotonic features \cite{roton1, roton2} and two-dimensional stable solitons \cite{2DSoliton}; while for a fermionic system the possibility to reach the Quantum Hall regime has been predicted \cite{David}. The first element used to investigate dipolar quantum gases was chromium with a magnetic moment of $\mu_\text{Cr}=6 \, \mu_\text{B}$ \cite{CrBEC,CrBECParis}. The strongly dipolar regime was reached by using a Feshbach resonance \cite{StrongDipol}. Recently a magneto-optical trap (MOT) for holmium atoms was realized \cite{HoMOT} and quantum degeneracy of both bosonic and fermionic erbium \cite{ErBEC,ErFermion} and dysprosium \cite{DyBEC,DyFermion} was achieved. These lanthanides provide even larger magnetic moments than chromium, in particular dysprosium being the most magnetic element with $\mu_\text{Dy}=10 \, \mu_\text{B}$.

In addition, these elements feature a rich atomic spectrum with several possible cooling transitions, hence, offering different cooling schemes. In the case of the present work with dysprosium, the strongest cycling transition at 421$\,$nm was used for Zeeman slowing and the MOT \cite{DyMOT}.
To reach suitable conditions to load into an optical dipole trap, the temperature was further decreased with a second stage MOT operating at the ultra narrow transition at 741$\,$nm with a natural linewidth of 1.8$\,$kHz \cite{Dy741}. 

In contrast, our approach is based on using the 421$\,$nm transition for Zeeman slowing and using the transition at 626 nm with a linewidth of 136(4)$\,$kHz \cite{626Line} for trapping the atoms with a narrow-line MOT, inspired by Er \cite{ErMOT} and Yb \cite{YbMOT} experiments. Based on this cooling scheme we capture up to $2.0\cdot10^8$ atoms at temperatures of 6$\, \mu$K. We achieve similar starting conditions, with the atoms already being in the energetically lowest Zeeman substate, compared to current work with ultracold dysprosium \cite{DyFeshbach}. These are ideal conditions for direct loading into an optical dipole trap. Preparing magnetic atoms in the lowest Zeeman substate is essential due to heating effects induced by dipolar relaxations \cite{drelax}.

The element dysprosium has two stable bosonic $^{162}\text{Dy} (26\%)$, $^{164}\text{Dy} (28\%)$ as well as two fermionic isotopes $^{161}\text{Dy} (19\%)$, $^{163}\text{Dy} (25\%)$ with high natural abundance. The electronic groundstate configuration $[\text{Xe}] 4\text{f}^{10} 6\text{s}^2$ offers an open 4f shell inside the closed 5s and 6s shells, which leads to a large orbital angular momentum of $L=6$. With its total electronic spin of $S=2$ the electronic groundstate has a total angular momentum of $J=8$, which is the origin of the high magnetic moment. In contrast to the bosonic isotopes, which have no nuclear spin, the fermionic isotopes possess a nuclear spin of $I=5/2$ leading to six additional hyperfine levels from $F=11/2$ to $F=21/2$ in the electronic groundstate.

The 421$\,$nm transition of dysprosium is used for Zeeman slowing and transverse cooling of the atomic dysprosium beam \cite{421Cool} and has a linewidth of $\Gamma_{\text{421}}/2\pi=32.2\,$MHz \cite{Dy741}, which leads to a Doppler temperature of $T_{\text{D,421}}=\hbar \Gamma / 2 k_\text{B}=773\, \mu$K. Its saturation intensity is $I_{\text{s,421}}= \pi h c \Gamma / 3 \lambda^3 =56.4\,$mW/$\text{cm}^2$. Magneto optical trapping is performed on the closed 626$\,$nm cycling transition. Due to its narrow linewidth of $\Gamma_{\text{626}}/2\pi=136\,$kHz the Doppler temperature results in $T_{\text{D,626}}=3.3\,\mu$K and the saturation intensity is $I_{\text{s,626}}=72\,\mu$W/$\text{cm}^2$. In the fermionic case the excited states of both transitions exhibit a hyperfine structure from $F'=13/2$ to $F'=23/2$ \cite{421HF,626HF}.

The 421$\,$nm laser light is produced by frequency doubling the infrared light of two Ti:sapphire laser systems. Both systems are locked together by a beat-note lock \cite{BeatNote}. The master Ti:sapphire laser is stabilized to an ultra low expansion cavity (ULE) and generates light for the transverse cooling and for absorption imaging. The slave system produces far red detuned light for the Zeeman slower (ZS). We create the orange 626$\,$nm light by sum frequency generation of a 1050$\,$nm and a 1550$\,$nm fiber laser in a periodically poled lithium niobate crystal (PPLN), inspired by \cite{626_sfg}. Both fiber lasers have a specified linewidth of $<\!10\,$kHz on timescales of 120$\,\mu$s. Frequency stabilization is achieved by coupling the orange light to the ULE cavity and using the feedback signal to lock the 1050$\,$nm fiber laser, while the 1550$\,$nm system is free running. We estimate a linewidth of $<\!30\,$kHz for the locked orange light and measured daily shifts of less than 20$\,$kHz. We employ an electro optical modulator (EOM) with a resonance frequency of 105$\,$kHz to broaden the linewidth of this laser to $\approx \! 70\, \Gamma_{\text{626}}$, which increases the capture velocity of the MOT. With this spectral broadener EOM we can achieve a higher atom number and a gain in the capture rate of our MOT.

\begin{figure}[t]
\centerline{\includegraphics[width=1\linewidth]{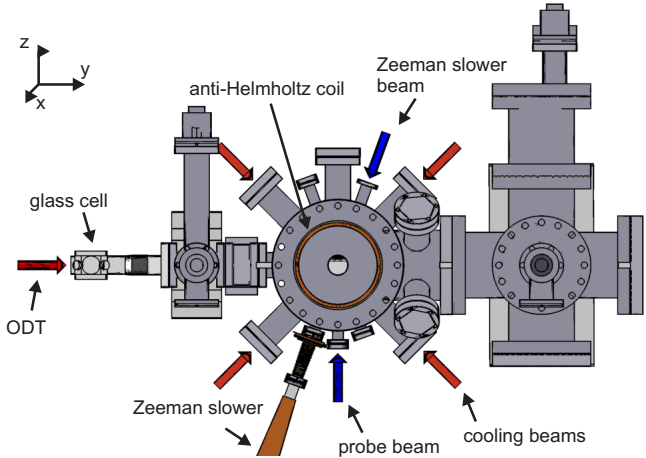}} 
\caption{(Color online) Schematic setup of our cold atom apparatus consisting of two main parts: a steel vacuum chamber for the MOT, and a glass cell with high optical access. The atomic beam is Zeeman slowed with 421$\,$nm light and travels with an angle of 20° to the gravitational direction $z$. The atoms are captured in the steal chamber, where the radial MOT beams are aligned under an angle of 45° in respect to the horizontal and vertical axis. The atoms are absorption imaged from below with 421$\,$nm light.}
\label{fig:setup}
\end{figure} 

A beam of dysprosium atoms is produced by a high temperature effusion cell at 1250°C. Before the atoms enter the ZS, a transverse cooling stage by a two dimensional optical molasses is provided. The beams are elliptically shaped ($w_\text{x}=6.8\,$mm, $w_\text{y}=1.7\,$mm) to achieve a higher spacial overlap with the atomic beam. We typically use a power of $P_{\text{trans}}=100\,$mW leading to an intensity of $I_{\text{trans}} \approx 9 \, I_{\text{s,421}}$ per beam, with a detuning of $-1 \, \Gamma_{\text{421}}$. Using transverse cooling we gain a factor of four in the atom number and increase the capture rate by a factor of five. In the spin-flip ZS the atoms are slowed down to a velocity of $v \approx 15\,$m/s. The ZS operates at a detuning of $-18 \, \Gamma_{\text{421}}$ and with a power $P_{\text{ZS}}=100\,$mW. The ZS beam is focused on the aperture of the effusion cell and has an estimated diameter of 18$\,$mm at the position of the MOT. This results in a light intensity of $I_{\text{ZS}} \approx 0.7\,I_{\text{s,421}}$. The effect of the ZS light on the MOT's lifetime is discussed in a later section.

The narrow-line MOT is set up in a retroreflected configuration where the radial beams are aligned under an angle of 45° with respect to the horizontal and vertical axis, as shown in Fig. \ref{fig:setup}. To accomplish a large trapping volume the beams have a diameter of 22.5$\,$mm and a light intensity of about $I_{\text{MOT}} = 370\,I_{\text{s,626}}$ per beam. The following presented data are focused on the $^{164}\text{Dy}$ bosonic isotope, while the other isotopes are qualitatively similar. 

We load more than $1.5\cdot10^8$ atoms with a temperature of about $500\,\mu$K  in 4$\,$s at a detuning of $\delta_\text{626}=-35\,\Gamma_\text{626}$ and with an axial magnetic field gradient of $\nabla B=3\,$G/cm. To reduce the temperature, and to increase the atomic number density, the MOT is compressed in 170$\,$ms by decreasing the MOT light intensity to $I_{\text{cMOT}}=0.15\, I_\text{{s,626}}$ and the detuning to $\delta_\text{{cMOT}}=-2.75 \,\Gamma_\text{{626}}$. To avoid atom losses during the compression stage we open the magnetic field gradient to $\nabla B=1.5\,$G/cm. Finally we end up with $1.5\cdot10^8$ atoms at a temperature of about $6\,\mu$K and an atomic number density of $8.6\cdot10^{10}\,$cm$^\text{-3}$, resulting in a phase space density of about $1.5\cdot10^{-5}$.  Moreover, we could prove by state selective absorption imaging that most of the atoms are automatically pumped by the MOT light to the lowest Zeeman state $m_\text{j}=-8$.

\begin{figure}[t]
\centerline{\includegraphics[width=1\linewidth]{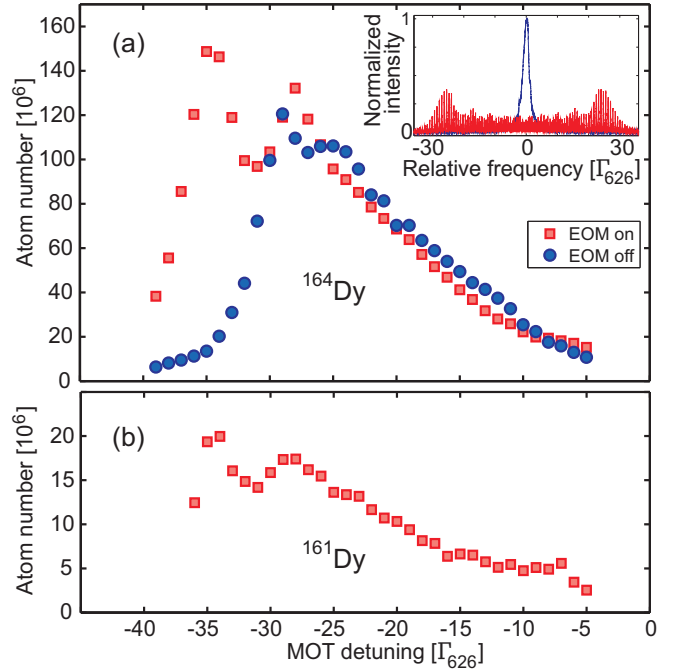}} 
\caption{(Color online) (a) Atom number as a function of MOT detuning for the $^{164}\text{Dy}$ boson. Spectral broadening of the 626 frequency allows us to reach higher detuning and consequently a higher atom number $N=1.5\cdot10^8$ at $\delta_\text{626}=-35\,\Gamma_\text{626}$ (squares). Without the spectral broadener only an atom number of $N=1.2\cdot10^8$ at $\delta_\text{626}=-29\,\Gamma_\text{626}$ is reached (circles). (inset) The EOM broadens the orange frequency up to $\approx \! 70 \, \Gamma_{\text{626}}$, which is shown in a transmission signal of the ULE cavity. (b) The $^{161}\text{Dy}$ fermion reaches a maximal atom number of $N=2.1\cdot10^7$ due to the fact that only the $F=21/2$ hyperfine state is trappable by the repumperless MOT as it also has a lower natural abundance.}
\label{fig:detuning}
\end{figure}

Fig. \ref{fig:detuning} shows the atom number as a function of the MOT detuning. Similar to the erbium \cite{ErMOT} and strontium \cite{SrMOT} narrow-line MOT gravity has a strong influence on the position and the shape of the atomic cloud. For higher detunings the cloud shifts downwards in the direction of gravity and increases its volume. Increasing the detuning the atom number increases due to an enlargement in the capture volume. After reaching its maximum the atom number decreases rapidly caused by the finite size of the trapping beams. Besides the increase in atom number and capture rate, the spectral broadener shifts the observed maximum in atom number towards a higher detuning. At a detuning of $\delta_\text{626}=-35\,\Gamma_\text{626}$ it provides frequency components from 0$\, \Gamma_\text{626}$ to -70$\, \Gamma_\text{626}$, resulting in a maximal atom number of up to $N=1.5\cdot10^8$ atoms. Between -28$\, \Gamma_{626}$ and -35$\, \Gamma_{626}$ the graph shows a drop in the atom number. This is caused by stray magnetic fields restricting the cloud's elongation. 

\begin{figure}[t]
\centerline{\includegraphics[width=1\linewidth]{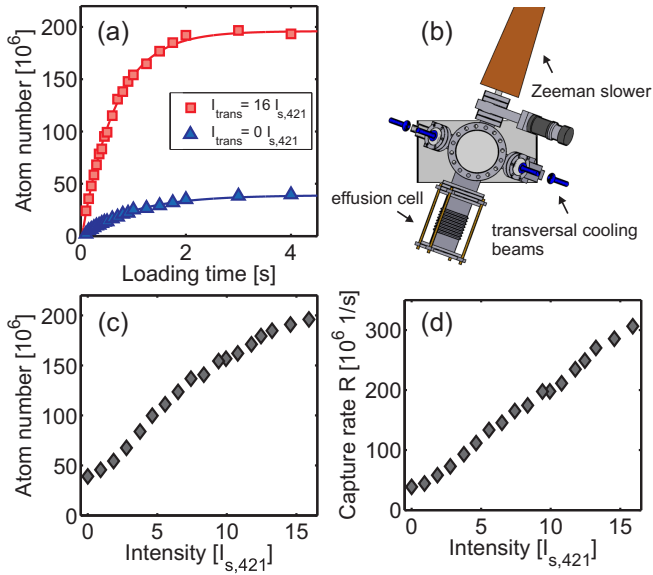}} 
\caption{(Color online) (a) Atom number as a function of the loading time at $\delta_\text{626}=-35\,\Gamma_\text{626}$. By fitting the standard loading rate equation (solid lines) one obtains a capture rate $R=3.1\cdot10^8\, 1/\text{s}$ and a decay rate $\gamma=1.6\,$1/s for $I_\text{trans}=16\,I_{\text{s,421}}$ (circles) and accordingly for no transverse cooling (triangles) $R=4.0\cdot10^7\, 1/\text{s}$ and $\gamma=1\,$1/s. (b) The transverse cooling is provided right after the oven aperture by shining in 421 nm light detuned by one linewidth. Atom number (c) and capture rate (d) are plotted as a function of the transverse cooling light intensity. The capture rate is increased by a factor of eight and the maximum atom number by a factor of five due to the transverse cooling.}
\label{fig:loading}
\end{figure} 

A transverse cooling stage has a major effect on the loading behaviour of our MOT, which is shown in Fig. \ref{fig:loading}. When transverse cooling before the Zeeman slower is employed the atom number reaches its steady state value $N_{\text{ss}}$ after around 3$\,$s. By fitting the standard loading rate equation $N(t)=N_{\text{ss}}(1-e^{-\gamma t})$ with $N_{\text{ss}}=R/\gamma$ we obtain a capture rate of $R=3.1\cdot10^8\,$1/s and a decay rate of $\gamma=1.6\,$1/s for the maximum available transverse cooling power of $P_{\text{trans}}=165\,$mW ($I=16\,I_{\text{s,421}}$) per beam. Using this intensity, we can increase the capture rate by a factor of eight and the maximum atom number by a factor of five to $2.0\cdot10^8$ atoms. 
 
\begin{figure}[t]
\centerline{\includegraphics[width=1\linewidth]{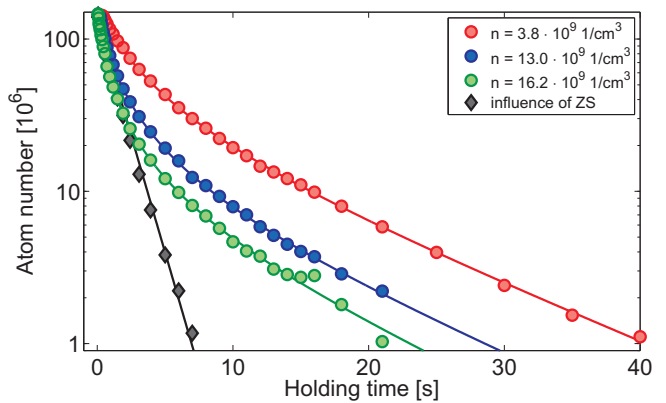}} 
\caption{(Color online) Atom number as a function of the holding time at a detuning $\delta_\text{626}=-35\,\Gamma_\text{626}$. The fast decay is caused by the ZS light (diamonds). A single exponential function is used to extract the time constant ($\gamma=1.4\,$1/s). In the case of no ZS light (circles) equation (1) was used to extract the two-body loss rate (solid lines). After the first fast two-body decay, we observe life times of around 12$\,$s, still limited by one-body scattering processes of the 626 nm light.}
\label{fig:lifetime}
\end{figure} 

We investigated the decay rate of the MOT concerning atom number density and ZS light by measuring the lifetime of our MOT. Here, after loading the MOT for 4$\,$s, the atomic beam was blocked. We observed that the maximum atom number is mainly limited by losses caused by the ZS light, as can be seen in Fig. \ref{fig:lifetime}. We attributed the fast decay ($\gamma=1.4\,$1/s) of the atom number to off-resonant pumping of the ZS light to excited states. This effect was present for any detuning $\delta_{\text{626}}$ meaning that a spatial separation of the atomic cloud from the ZS light, as reported in \cite{ErMOT}, seems not possible with our configuration, where the ZS light propagates at an angle of 20° with respect to gravitational direction.

As a consequence, we switched off the ZS light and the ZS field resulting in a longer lifetime of about 12$\,$s. We also turned off the spectral broadener as it did not affect the lifetime. Note that the atomic cloud changes its position by missing ZS light pressure and ZS magnetic field. We still observe a fast decay at the beginning which we attribute to two-body losses due to light induced collisions. To extract the two-body loss rate $\beta$ we use the following equation for the evolution of the atom number \cite{twobodyformular}:

\begin{equation} 
N(t)=\frac{N_{\text{ss}} \gamma e^{-\gamma t}}{\gamma+\frac{\beta N_{\text{ss}} \rho_{\text{ee}}(1-\rho_{\text{ee}})}{\overline{V}}(1-e^{-\gamma t})},
\end{equation}
where $N_{\text{ss}}$ is the steady state atom number, $\gamma$ the one-body loss rate, $\rho_{\text{ee}}$ the relative excited state population and $\overline{V}=2\sqrt{2}\pi^{3/2}\sigma_\text{x}\sigma_\text{y}^2$ the effective volume with $\sigma_\text{x}$ the axial and $\sigma_\text{y}$ the radial cloud size. We took lifetime curves for three different atomic number densities $n = N_{\text{ss}}/\overline{V}$ by changing the trapping volume with the magnetic field gradient (Fig. \ref{fig:lifetime}). In addition we investigated the decay rate of the compressed MOT where we have a ten times lower relative excited state population due to the smaller detuning and lower light intensity. As a result we estimate the two-body loss rate to $\beta=3.0(3)\cdot10^{-9}\,\text{cm}^3/\text{s}$, independent of the atomic number density and the relative excited state population of $\rho_{\text{ee}}=0.037$.

The slow decay at longer times is caused by pumping processes of the orange light. To investigate this we reduced the intensity after the loading phase to different intensities between $1\,I_{\text{s,626}} \leq I_{626} \leq 18\,I_{\text{s,626}}$ and at low intensities were able to achieve lifetimes up to 35$\,$s.

Futhermore, we achieved a MOT for the bosonic $^{162}\text{Dy}$ isotope with $N=1.3\cdot10^8$ atoms at a temperature of $6\, \mu$K. In comparison to the $^{164}\text{Dy}$ isotope the lower atom number is caused by the lower natural abundance. In the fermionic case we were able to trap up to $N=2.1\cdot10^7$ $^{161}\text{Dy}$ atoms without additional repumpers, see Fig. \ref{fig:detuning}. Taking into account the natural abundance and the fact that only the $F=21/2$ hyperfine state is trappable by the MOT, the reached atom number is higher than expected. This means the repumperless ZS light pumps a small amount of atoms from lower hyperfine states to the $F=21/2$ state. For all isotopes we could see the same effects of the transverse cooling and the spectral broadener. 

In summary we have realized a narrow-line MOT using the 626$\,$nm transition in dysprosium. With this method we trap up to  $N=2.0\cdot10^8$ atoms at a temperature of $6\,\mu$K. Our scheme is a simple and effective way to provide cold atoms for loading directly into an optical dipole trap. We currently load $N=20\cdot10^6$ atoms in our single beam dipole trap. After loading we transport the atoms from the MOT chamber over a distance of 40$\,$cm to the glass cell by moving the last focusing lens of the dipole trap, which is mounted on an air bearing stage. We are able to transport $N=11\cdot10^6$ to the glass cell, where no extra loss occurs as compared to keeping the atoms in the dipole trap in the steel chamber. The transport efficency is hence only limited by the normal background loss during the time needed to move the lens.
%

This work is supported by the German Research Foundation (DFG) within SFB/TRR21. H.K. acknowledges support by the 'Studienstiftung des deutschen Volkes'.



\begin{thebibliography}{10}
\newcommand{\enquote}[1]{``#1''}

\bibitem{Review}
T.~Lahaye, C.~Menotti, L.~Santos, M.~Lewenstein, and T.~Pfau, Reports on
  Progress in Physics \textbf{72}, 126401 (2009).

\bibitem{roton1}
L.~Santos, G.~V. Shlyapnikov, and M.~Lewenstein, Phys. Rev. Lett. \textbf{90},
  250403 (2003).

\bibitem{roton2}
S.~Ronen, D.~C.~E. Bortolotti, and J.~L. Bohn, Phys. Rev. Lett. \textbf{98},
  030406 (2007).

\bibitem{2DSoliton}
I.~Tikhonenkov, B.~A. Malomed, and A.~Vardi, Phys. Rev. Lett. \textbf{100},
  090406 (2008).

\bibitem{David}
D.~Peter, A.~Griesmaier, T.~Pfau, and H.~P. B\"uchler, Phys. Rev. Lett.
  \textbf{110}, 145303 (2013).

\bibitem{CrBEC}
A.~Griesmaier, J.~Werner, S.~Hensler, J.~Stuhler, and T.~Pfau, Phys. Rev. Lett.
  \textbf{94}, 160401 (2005).

\bibitem{CrBECParis}
Q.~Beaufils, R.~Chicireanu, T.~Zanon, B.~Laburthe-Tolra, E.~Mar\"echal,
  L.~Vernac, J.-C. Keller, and O.~Gorceix, Phys. Rev. A \textbf{77}, 061601
  (2008).

\bibitem{StrongDipol}
T.~Lahaye, T.~Koch, B.~Fr\"ohlich, M.~Fattori, J.~Metz, A.~Griesmaier,
  S.~Giovanazzi, and T.~Pfau, Nature \textbf{448}, 672 (2007).

\bibitem{HoMOT}
J.~Miao, J.~Hostetter, G.~Stratis, and M.~Saffman, arXiv:1401.4156  (2014).

\bibitem{ErBEC}
K.~Aikawa, A.~Frisch, M.~Mark, S.~Baier, A.~Rietzler, R.~Grimm, and
  F.~Ferlaino, Phys. Rev. Lett. \textbf{108}, 210401 (2012).

\bibitem{ErFermion}
K.~Aikawa, A.~Frisch, M.~Mark, S.~Baier, R.~Grimm, and F.~Ferlaino, Phys. Rev.
  Lett. \textbf{112}, 010404 (2014).

\bibitem{DyBEC}
M.~Lu, N.~Q. Burdick, S.~H. Youn, and B.~L. Lev, Phys. Rev. Lett. \textbf{107},
  190401 (2011).

\bibitem{DyFermion}
M.~Lu, N.~Q. Burdick, and B.~L. Lev, Phys. Rev. Lett. \textbf{108}, 215301
  (2012).

\bibitem{DyMOT}
M.~Lu, S.~H. Youn, and B.~L. Lev, Phys. Rev. Lett. \textbf{104}, 063001 (2010).

\bibitem{Dy741}
M.~Lu, S.~H. Youn, and B.~L. Lev, Phys. Rev. A \textbf{83}, 012510 (2011).

\bibitem{626Line}
M.~Gustavsson, H.~Lundberg, L.~Nilsson, and S.~Svanberg, J. Opt. Soc. Am.
  \textbf{69}, 984 (1979).

\bibitem{ErMOT}
A.~Frisch, K.~Aikawa, M.~Mark, A.~Rietzler, J.~Schindler,
  E.~Zupani\ifmmode~\check{c}\else \v{c}\fi{}, R.~Grimm, and F.~Ferlaino, Phys.
  Rev. A \textbf{85}, 051401 (2012).

\bibitem{YbMOT}
T.~Kuwamoto, K.~Honda, Y.~Takahashi, and T.~Yabuzaki, Phys. Rev. A \textbf{60},
  R745 (1999).

\bibitem{DyFeshbach}
K.~Baumann, N.~Q. Burdick, M.~Lu, and B.~L. Lev, Phys. Rev. A \textbf{89},
  020701 (2014).

\bibitem{drelax}
S.~Hensler, J.~Werner, A.~Griesmaier, P.~Schmidt, A.~G\"{o}rlitz, T.~Pfau,
  S.~Giovanazzi, and K.~Rza\ifmmode \mbox{\c{}}\else
  \c{}\fi{}\ifmmode~\dot{z}\else \.{z}\fi{}ewski, Applied Physics B
  \textbf{77}, 765 (2003).

\bibitem{421Cool}
N.~Leefer, A.~Cing\"oz, B.~Gerber-Siff, A.~Sharma, J.~R. Torgerson, and
  D.~Budker, Phys. Rev. A \textbf{81}, 043427 (2010).

\bibitem{421HF}
N.~Leefer, A.~Cing\"{o}z, and D.~Budker, Opt. Lett. \textbf{34}, 2548 (2009).

\bibitem{626HF}
W.~Hogervorst, G.~Zaal, J.~Bouma, and J.~Blok, Physics Letters A \textbf{65},
  220  (1978).

\bibitem{BeatNote}
U.~Sch\"{o}nemann, H.~Engler, R.~Grimm, M.~Weidem\"{u}ller, and
  M.~Zielonkowski, Review of Scientific Instruments \textbf{70}, 242 (1999).

\bibitem{626_sfg}
A.~Wilson, C.~Ospelkaus, A.~VanDevender, J.~Mlynek, K.~Brown, D.~Leibfried, and
  D.~Wineland, Applied Physics B \textbf{105}, 741 (2011).

\bibitem{SrMOT}
H.~Katori, T.~Ido, Y.~Isoya, and M.~Kuwata-Gonokami, Phys. Rev. Lett.
  \textbf{82}, 1116 (1999).

\bibitem{twobodyformular}
M.~Prentiss, E.~L. Raab, D.~E. Pritchard, A.~Cable, J.~E. Bjorkholm, and
  S.~Chu, Opt. Lett. \textbf{13}, 452 (1988).

\end{thebibliography}
\end{document}